\documentclass[preprint]{aa}
\usepackage{graphicx}
\usepackage{natbib}
\usepackage{gensymb}

\begin{document}
\title{The dynamical behaviour of a jet in an on-disk coronal hole observed with AIA/SDO} 

\author{K.~Chandrashekhar\inst{1}
R. J. Morton\inst{2}
        D.~Banerjee\inst{1}
        \and
        G. R. Gupta\inst{3}}
\institute{Indian Institute of Astrophysics, Koramangala, Bangalore 560034.
\and Mathematics and Information Science, Pandon Building, Camden Street, Northumbria University, 
Newcastle Upon Tyne, NE1 8ST, UK
\and
Max-Plank-Institut f\"{u}r Sonnensystemforschung (MPS), 37191, Katlenburg-Lindau, Germany}

\date{\today}  
 
 \abstract
 {} 
{EUV jets situated in coronal holes are thought to play an important role in supplying heated material to 
the corona and solar wind. The multi-wavelength capabilities and high signal-to-noise of detectors 
on-board SDO allows for detailed study of these jet's evolution. We aim to exploit SDO's capabilities to 
reveal information on the jet dynamics and obtain estimates for plasma properties associated with the 
jet.}
{We study the dynamics an EUV jet with AIA/SDO at a coronal hole boundary. The details of the jet 
evolution are discussed and measurements of the jet's parameters, e.g. length, width, life time, outward 
speed, are obtained. Further, automated emission 
measure analysis is exploited to determine estimates for the temperature and density of the jet. A 
propagating transverse wave supported by the jet spire is also observed. Measurement of the wave 
properties are exploited for magneto-seismology and are used in conjunction with the emission measure 
results to estimate the magnetic field strength of the jet.}
{We present a detailed description of the jet's evolution, with new evidence for plasma flows, prior to the 
jet's initiation, along the loops at the base of the jet and also find further evidence that flows along the 
jet spire consist of multiple, quasi-periodic small-scale plasma ejection events. In addition, DEM analysis 
suggests that the jet has temperatures of $\log{5.89\pm0.08}$~K and electron densities of 
$\log{8.75\pm0.05}$~cm$^{-3}$. Measured properties of the transverse wave 
suggest the wave is heavily damped as it propagates along the jet spire with speeds of $\sim110$~km/s. 
The magneto-seismological inversion of the wave parameters provides values of $B=1.21\pm0.2$~G along the jet spire, which is in line with previous estimates for open fields in coronal holes. }
{}


\titlerunning{Jet dynamics}
\authorrunning{K. Chandrashekhar et al.}

\maketitle

\section{Introduction}
\label{int}

Over the last two decades, it has become evident that the large-scale ejection of collimated plasma from 
the lower solar atmosphere in to the corona occurs on a regular basis (e.g., \citealp{SHIbetal1992}; 
\citealp{SHIMetal1996}; \citealp{2007Sci...318.1580C}). These events, referred to as EUV, X-ray or 
coronal plasma jets, are common in coronal holes and at the coronal hole boundaries 
(\citealp{SAVetal2007}). These jets are often considered as a potential mechanism for supplying the 
corona with heated plasma and maintaining the solar wind (e.g., \citealp{SUBetal2010}; 
\citealp{MOOetal2011}).

Recent high-resolution and high-cadence multi-wavelength observations (e.g., using Hinode, STEREO, 
Solar Dynamic Observatory - SDO) have allowed for detailed study of coronal hole jets, providing 
information on their inherent dynamic behaviour (e.g., \citealp{KAMetal2007}; 
\citealp{PATetal2008}; \citealp{NISetal2009}; \citealp{FILetal2009} \citealp{LIUetal2009, LIUetal2011}; 
\citealp{KAMetal2010}; \citealp{SHEetal2011}; 
\citealp{YANetal2011}; \citealp{MORetal2012b}; \citealp{CHEetal2012}) and estimates for the plasma 
parameters (\citealp{CULetal2007}; \citealp{DOSetal2010}; \citealp{HEetal2010}; \citealp{MAD2011}; 
\citealp{NISetal2011}). In brief, the observations tend to show the jets as an inverted-Y shape, consisting 
of a collimated ejection that forms the spire and with a number of heated loop structures seen at the 
base. Measured plasma outflow velocities along the spire display a range of speeds, some showing 
impulsive (and occasionally quasi-periodic) phases with Alfv\'enic speeds (200-600~km/s) and others 
with continuous outflow velocities on the order of the sound speed (30-100~km/s). The jets also seem to 
demonstrate two distinct sub-classes of behaviour allowing observations to be separated into standard 
or blow-out jet categories (e.g., \citealp{MOOetal2010}). The standard jets consist of hot EUV and soft X-
ray components, while the blow-out jets also have the additional emission of a cool plasma component 
(sometimes referred to as the plasma curtain) that is visible in spectral lines that typically correspond to 
chromospheric and Transition Region emission.

In conjunction with the observations, advanced numerical modelling of jet formation has also been 
undertaken (e.g., \citealp{MORetal2008}; \citealp{PARetal2009}; \citealp{MORGAL2013}).  The formation mechanism of the 
coronal jets is typically assumed to be due to reconnection of emerging flux with pre-existing flux 
\citep{1995Natur.375...42Y}. Magnetic energy release(s) due to the reconnection is converted to either: 
the heating the plasma; wave excitation; kinetic energy (i.e., bulk plasma motions). The simulations of 
the emerging flux scenario seem to provide a good qualitative agreement with observations.

As mentioned, magnetohydrodynamic (MHD) waves are excited during magnetic reconnection and are an 
effective mechanism for the transport of energy. This feature is highlighted in observations 
(\citealp{KAMetal2010}; \citealp{SHEetal2011}; \citealp{MORetal2012b}; \citealp{CHEetal2012}) and 
simulations (\citealp{PARetal2009}), where non-linear torsional motions are seen to drive magnetic field 
and plasma away from the reconnection site. The presence of MHD waves in and around magnetic null-
points can also lead to periodic reconnection (e.g., \citealp{MURetal2009}; 
\citealp{MCLetal2009,MCLetal2012b,MCLetal2012}; \citealp{HEGetal2009}). Evidence for periodic 
reconnection has been noted by \cite{MORetal2012b}, who 
observe the periodic ($\sim50$~s) ejection of small-scale jets that appear to supply the main jet spire 
with heated plasma. The presence of wave phenomenon in these jet events provides a unique opportunity 
for magneto-seismology, i.e., the inversion of measured wave parameters to reveal information on the 
local magnetised plasma. The potential for such studies has been demonstrated in 
\citealp{2012ApJ...744....5M}, where the plasma temperature, magnetic field gradients and plasma 
density gradients along a cool loop, that erupted from the base arcade of the jet, are determined.

 {Magneto-seismology has previously been used to infer magnetic field strengths and 
gradients in a number of other magnetic structures, e.g., coronal loops (recently, e.g., \citealp{VERetal2013b}; or for reviews see, e.g., 
\citealp{RUDERD2009}; \citealp{ANDetal2009}; \citealp{ERDGOO2011}), EUV jets (\citealp{2012ApJ...744....5M}); 
spicules (\citealp{VERTetal2011}).} However, the real key to determining the plasma parameters will be a combination of spectroscopic 
techniques with magneto-seismology. In this paper, we study the properties of a coronal jet using data 
from multiple channels of the Atmospheric Imaging Assembly (AIA) on-board SDO.  The temperature and 
density of the jet are determined using differential emission measure (DEM) analysis and 
provide comparable results to previous measurements. Further, the presence of 
transverse perturbations of the jets spire are also observed and measurements of the waves amplitude, 
period and phase speed are made. We then demonstrate the diagnostic potential from the union of 
spectroscopic and magneto-seismological techniques by using the results to estimate the magnetic field 
in the jets spire.

\section{Observations}
\label{obs}
SDO is the first mission of NASA's Living With a Star Program. It 
was launched on 11 February 2010, and allows nearly continuous observations of the Sun. The AIA 
instrument on-board SDO provides an unprecedented view of the solar corona at a high cadence in 
multiple EUV wavelengths nearly simultaneously. The high spatial and temporal 
resolutions of AIA are 0.6~arcsec per pixel and 12~seconds respectively \citep{2011SoPh..tmp..172L} and 
this allows the detailed study of transient solar events. AIA provides 
narrow-band imaging in seven EUV bandpasses which are centered on \ion{Fe}{xviii} (94~${\AA}$), 
\ion{Fe}{vii,xxi} (131~${\AA}$), \ion{Fe}{ix} (171~${\AA}$), \ion{Fe}{xii,xxiv} (193~${\AA}$), \ion{Fe}{xiv} 
(211~${\AA}$), \ion{He}{ii} (304~${\AA}$),  and \ion{Fe}{xvi} (335~${\AA}$) spectral lines. 

\begin{figure}
\centering
\includegraphics[scale=0.38, clip=true, viewport=0.0cm 0.0cm 20.cm 20.cm]{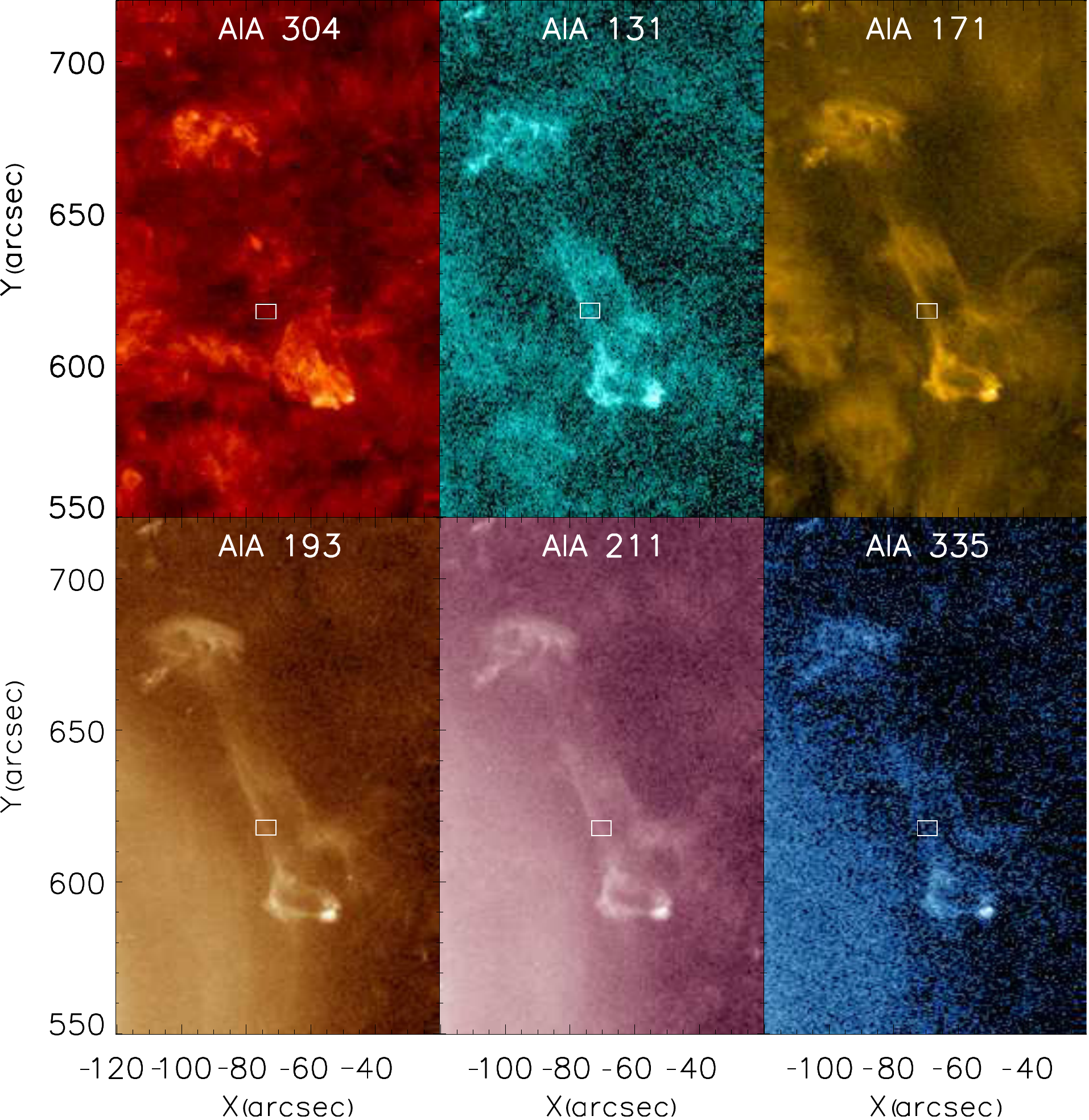}
\includegraphics[scale=0.7, clip=true, viewport=0.0cm 0.0cm 12.cm 10.6cm]{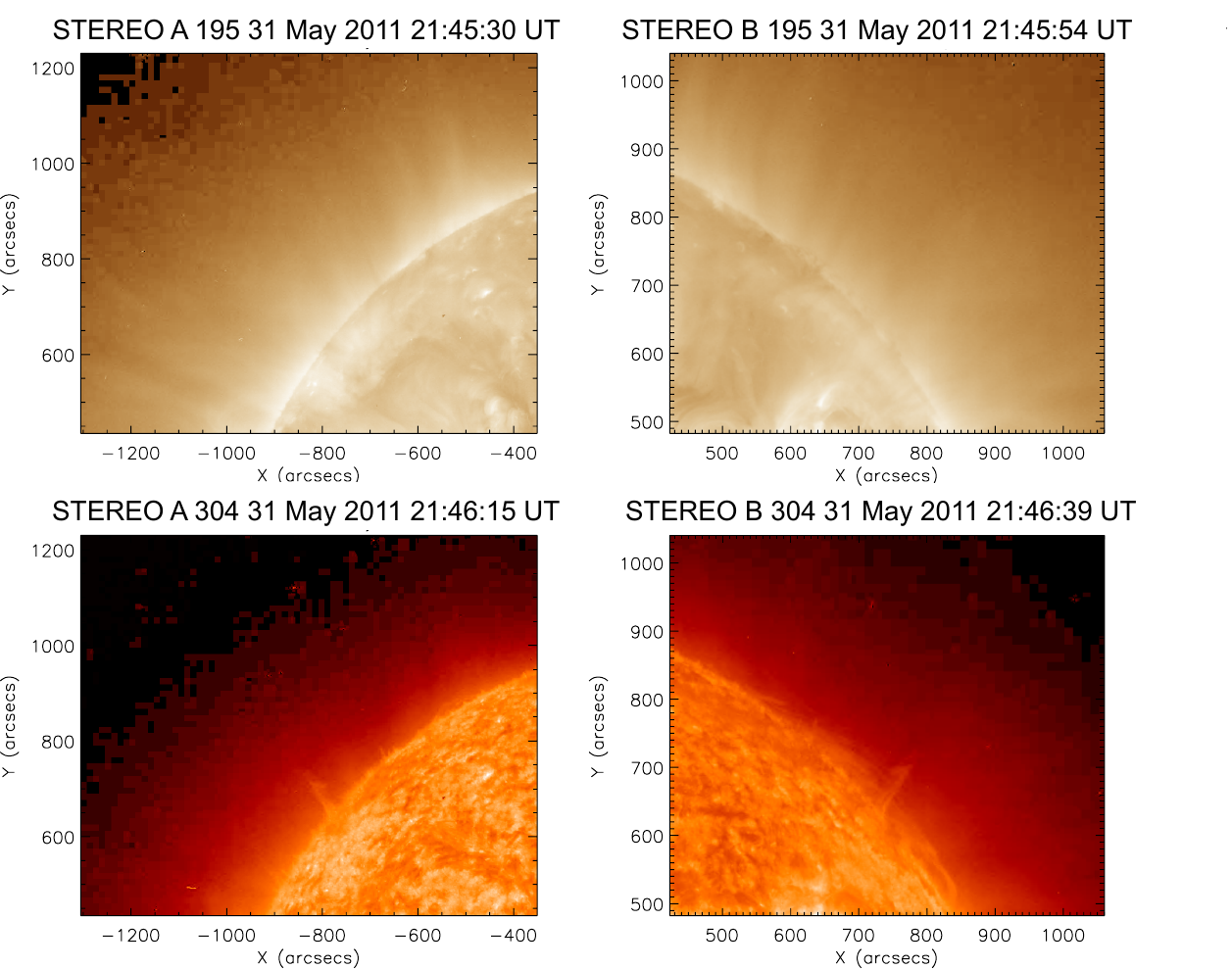}
\caption{The top panels display SDO/AIA images of the polar coronal hole jet. The wavelength 
corresponding to each channel is given in the panels. The rectangular box in each frame marks the 
location of the pixels used to generate the light curves in Fig~\ref{aia-lc}. The bottom panels are STEREO/SECCHI EUVI images of the 
same jet. The position of the jet is not obvious in the 195~{\AA} bandpass but can be seen clearly above the limb in 304~{\AA}.}
\label{aiajet}
\end{figure}

For this study, we have used data taken with SDO during a period from 21:00 to 22:00~UT 31 May 2011. 
The data set is of a region containing the boundary of an on-disk coronal hole situated in the 
northern hemisphere. The level~1.0 data are processed to remove bad pixels, spikes due to radiation 
effects, and correct the image flat-field. Level~1.0 data are then converted to level~1.5 using {\sf 
aia\_prep.pro} procedure in {\sf Solar Soft} (SSW). This procedure also adjusts the different filter images 
to a common 0.6$\arcsec$ plate-scale. To correct for rotation and achieve sub-pixel alignment accuracy 
of the time-series, we apply cross-correlation techniques. Figure~\ref{aiajet} shows images of the jet at 
it's maximum activity in the different AIA channels. The boundary of the coronal hole can be identified
in the 193~{$\AA$}, 211~{$\AA$} and 94~{$\AA$}, where the closed corona is identified by the bright 
emission in the lower left hand corner.

The SDO data is supplemented with data from SECCHI/STEREO (\citealp{HOWetal2008}), using the 
304~{\AA} and 195~{\AA} channels. The STEREO data is processed using the standard software. The jet is 
only visible in the 304~{\AA} of both A and B space-crafts and it's signature is conspicuously absent in 
195~{\AA} images. The STEREO images (Fig.~\ref{aiajet}) reveal that the jet makes an angle almost 
normal to the surface, hence, is inclined to the projected view of SDO.
The angle between the jet and the SDO projection is $\sim36$~\degree, which is in line with the jet's 
latitude of 37\degree in SDO images.

\begin{figure*}
\centering
\includegraphics[scale=0.7]{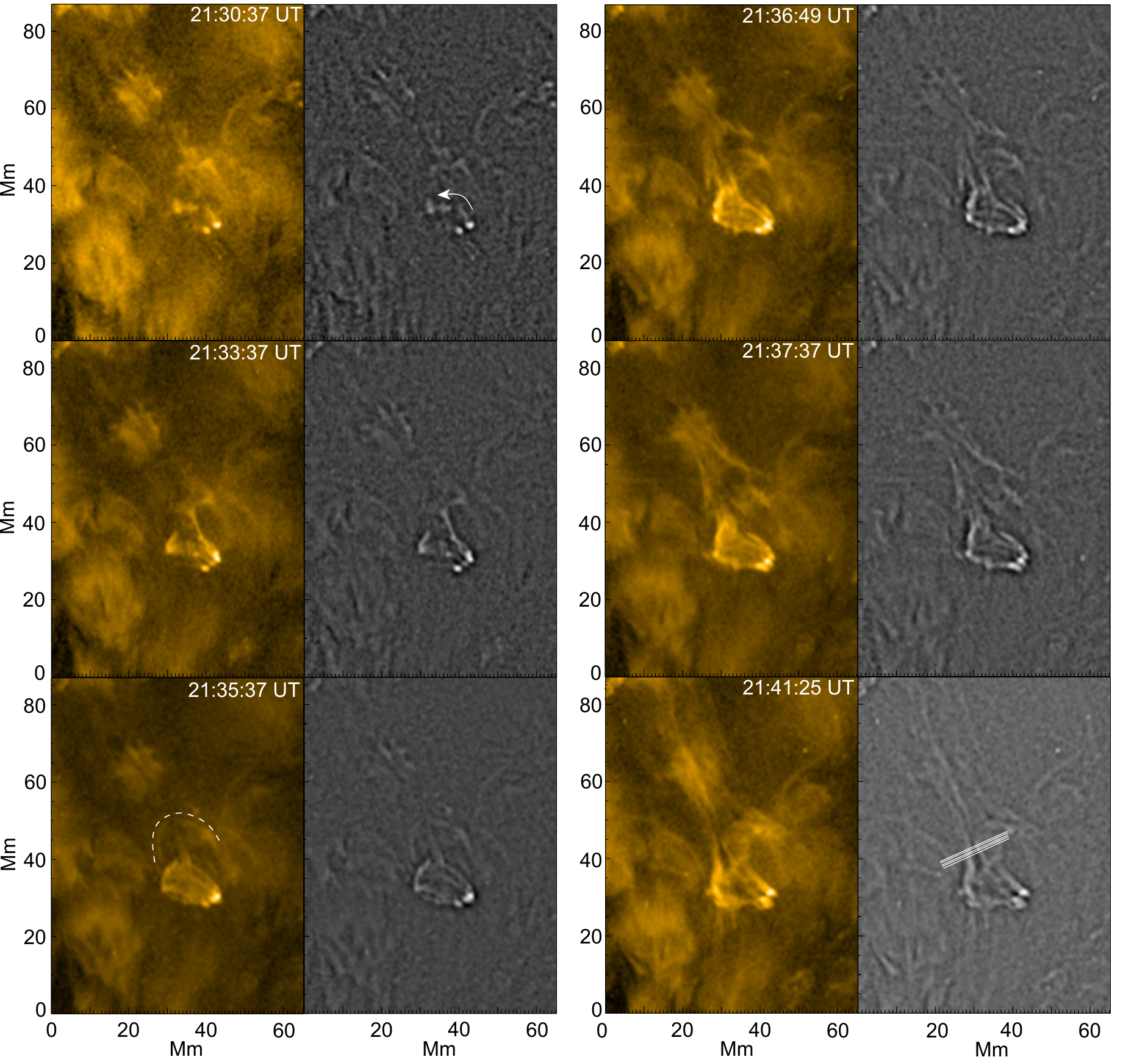}
\caption{Evolution of the jet in 171~{$\AA$}. The left hand panels show log intensity and the right hand 
panels the unsharp masked log intensities. Top left panels: Enhanced emission at arcade foot-points and 
observed outflow. Middle left: Apparent reconfiguration of the magnetic field and increased emission 
along the base arcade. Bottom left: Rise phase of a loop overlying the base arcade. Top right: Formation 
of jet spire and the rising loop apparently connects with unseen open-field. Middle right: Initiation of 
out-flowing plasma along the jet spire. Bottom right: Elongated jet structure with over-plotted cross-cuts 
used for anlaysis of the kink motion.}
\label{jet_evol}
\end{figure*}

\section{Dynamical evolution of the jet}
\label{sec:jet}
Now, we describe the jet evolution and highlight some of the more interesting aspects. The onset of 
dynamic behaviour begins with the appearance of a pair of bright points (Fig.~\ref{jet_evol} first panel), 
which mark the foot-points of a loop arcade (the arcade becomes conspicuous at a later time). 

On viewing co-temporal magneto-grams taken with HMI, the region shows only weak magnetic flux 
elements present in this region. A popular mechanism for the initiation of coronal jets is the interaction 
between an emerging bipolar region of magnetic flux and existing unipolar field in the coronal holes 
\citep{1995Natur.375...42Y}. The emergence of a bipolar region or the injection of twisted flux into an 
existing bipolar region eventually leads to reconnection between the bipolar and unipolar fields. 
The weak nature of the signal in the magnetograms means we cannot use them to confirm such a 
hypothesis.  

The injection of new magnetic flux into the region can lead to the build-up of currents (e.g., 
\citealp{PARetal2009}), which could potentially be dissipated and heat the plasma in the lower 
atmosphere. We 
suggest that the observed brightenings are potential signatures of the injection of new flux into the 
region and the subsequent heating. The increase in pressure at the foot-points due to the heating would 
lead to a pressure gradient and drive flows of heated plasma away from the region. Such outflows from 
the bright-points are indeed visible (see, online movie 1 of Figure~\ref{jet_evol}) and the flow is 
highlighted in Figure~\ref{jet_evol} first panel. The flow of heated 
plasma highlights an existing loop structure and the the arrow demonstrates the flow direction. 

At approximately 250~s later, there is an apparent initial reconfiguration of the magnetic field, possibly 
reconnection events, where a loop structure(s)  begin to open up (Fig.~\ref{jet_evol} second panel). This 
is coupled with increased emission and the loop arcade becomes increasingly more visible. Drawing from 
the findings of simulations (e.g., \citealp{MORGAL2013}), the increased visibility of the loop arcade could 
be the signature that a current sheet has formed. This would occur if new flux has emerged and is 
impinging on the pre-existing field, and would also lead to reconnection between the two fields. The 
enhanced emission may come from plasma heating due to the dissipation of the currents or is related to 
the presence of dense, warm material.

It is possible to observe out-flowing plasma along the newly opened field lines. Co-temporal 304~{\AA}
images (online movies 2, 3 and 4) show that a significant amount of cool material is ejected at this stage, 
making up a portion of the out-flowing material. Again, with reference to simulations, the ejection of the 
cool, dense material can be associated with flux emergence. The emerging flux would bring dense 
material with it as it rises into the atmosphere. Reconnection of the emerging flux with the ambient, open 
field leads to dense material being ejected along the reconnected field lines.

In conjunction with the outflow, an extended low lying 
loop, almost perpendicular to the arcade, becomes visible. The loop structure appears to have one 
foot-point rooted in the left-hand side of the arcade and the other in a region at least 10~Mm from the 
arcade. In time, the loop expands upwards into the atmosphere, but not horizontally. The loop is 
deformed into a semi-elliptical shape during the expansion and the structure is highlighted by the 
dashed line in Fig.~\ref{jet_evol} panel three. It is unclear whether the loop begins to rise due to the 
upward pressure from the out-flowing plasma, however, when the loop becomes noticeable it appears at 
the leading edge of the outflow (clearly seen in 304~{\AA}) and continues to rise with the out-flowing 
plasma. The measured rise speed of the loop in the plane is 25$\pm$5 km\,s$^{-1}$, however, this 
value is likely subject to effects of projection and is in reality, larger than the measured value.

In Fig.~\ref{jet_evol} panel four, the loop is now clearly visible. Overlying open fields give the appearance 
of barbs in the direction perpendicular to the magnetic field of the loop. Careful study of the movie 
suggests these features are not related to the rising loop, they just become visible at a similar time. 
Observed movement of these features away from the arcade may point towards them being additional 
field lines that have reconnected. 
Just after this frame, the loop opens up in the middle and connects to non-visible, presumably 
pre-existing field, forming a funnel shape (Fig.~\ref{jet_evol} panel five) which eventually becomes the 
main spire of the jet (Fig.~\ref{jet_evol} panel six). Soon after the funnel is formed, the ejection of plasma 
begins and there is visible evidence for torsional/helical motions during the first phase of ejection (see, 
Fig.~\ref{fig:wave} and online movie 1).

\begin{figure}
\centering
\includegraphics[width=9cm]{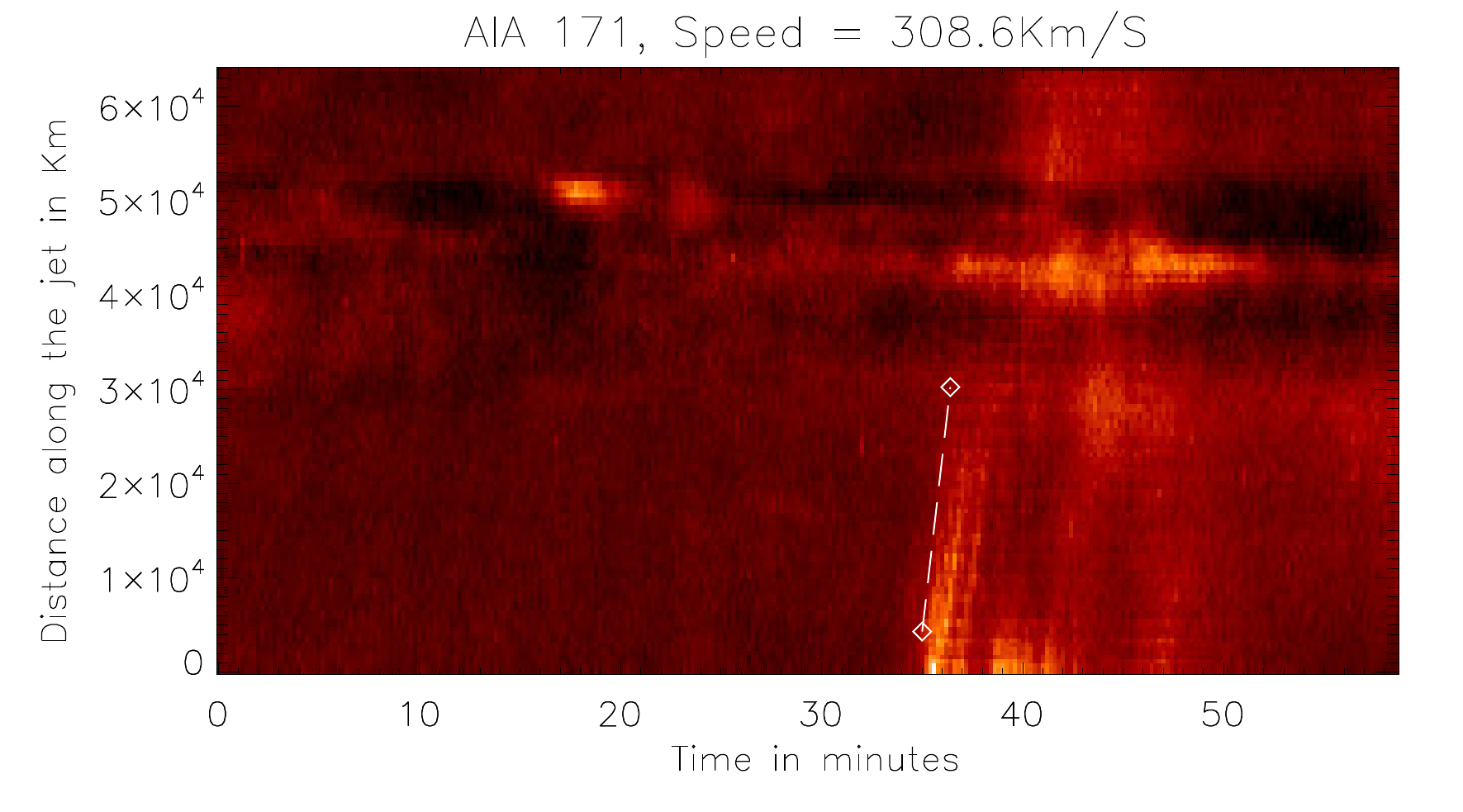}
\caption{ Time-distance diagrams for the jet as seen by the 171~{\AA} channel. The diamond symbols 
indicate the two points selected for producing the slope and dashed line indicates the fit obtained. }
\label{aia-xt}
\end{figure}

 Plasma is seen to flow out along the spire for at least 10 minutes, with the intensity of the out-flowing 
plasma decreasing over time. The outward speed of the plasma is estimated in the different channels by 
plotting time-distance maps. The time-distance maps are produced by plotting intensity along the length 
of the jet versus time, with the slope of the line fitted to the bright region giving the speed of the jet. 
Fitting for the time-distance map is done as follows. First, we select two points well separated along the 
left side of the bright emission in the time-distance map. From time-distance map data we produce 
light-curves at the various heights along the jet spire. Peaks in the light-curves are assumed to 
correspond to the time at which out-flowing plasma reaches each particular height. Figure~\ref{aia-xt} shows time-distance maps for the jet as seen in 171~{\AA} channel.  The 
diamond symbols indicate the two points selected for 
producing the light-curves and dashed line indicates the fit obtained.  The speeds as measured by the 
slopes corresponding to different channels are as follows: AIA 131~{\AA} (334 km\,s$^{-1}$), AIA 
171~{\AA} (308 km s$^{-1}$), AIA 193~{\AA} (360  km s$^{-1}$), AIA 211~{\AA} 
(312~km\,s$^{-1}$) and AIA 335~{\AA} (312~km\,s$^{-1}$). The errors in the estimate of speeds are 
$\pm 36$~km\,s$^{-1}$ guided by the pixel size and cadence of observation. However, as noted the jet 
is at an angle to the AIA projection, so the true speeds are given by dividing by $\cos\theta$, hence the 
corrected outflow speeds are 412~km\,s$^{-1}$, 381~km\,s$^{-1}$, 445~km\,s$^{-1}$, 
386~km\,s$^{-1}$, 386~km\,s$^{-1}$.

\begin{figure}[!t]
\centering
\includegraphics[width=8cm]{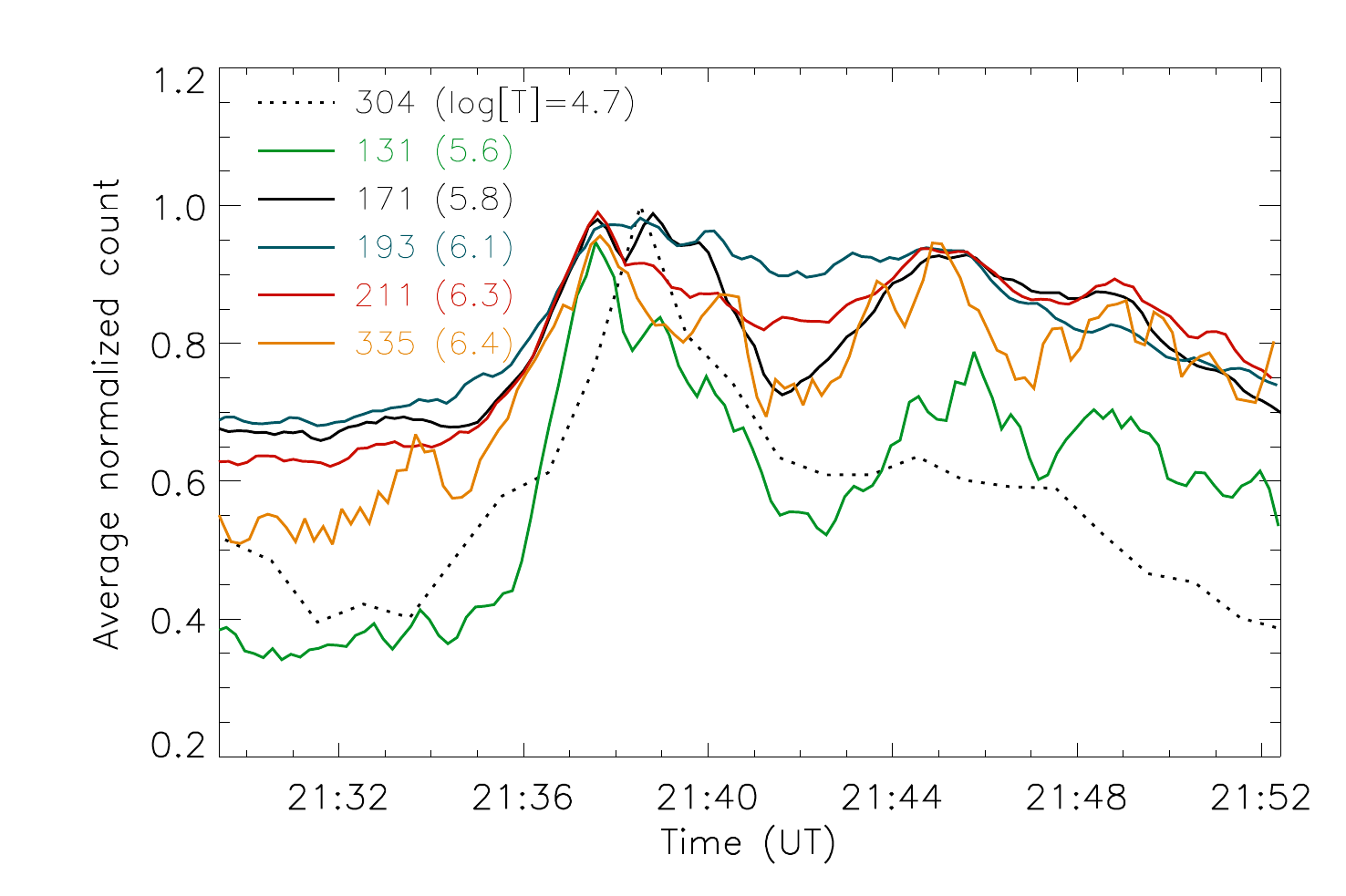}
\caption{Light-curves corresponding to the rectangular square box region as shown in Fig.~\ref{aiajet}. 
Different colors correspond to the different channels as labelled in the figure, along with the peak 
formation temperatures of the dominant ions in parenthesis. \label{aia-lc} }
\end{figure}

Figure~\ref{aia-lc} shows the light curves of the different channels in AIA corresponding to a small region 
shown as square box in Figure~\ref{aiajet}. The initial eruptive phase of the jet is seen to begin
around 21:37 UT. Further, additional enhancements in emission occur at 21:45 UT, except in the 
304~{\AA} line, suggesting a second eruptive stage with the emission of hot plasma only. The 131~{\AA} 
and 335~{\AA} display clear evidence for multiple peaks. This behaviour is not as evident in the other 
light curves but it can clearly be seen from Figure~\ref{aia-xt} that there are multiple outflows events. 

\begin{figure}
\centering
\includegraphics[scale=.55,clip=true, viewport= 0.0cm 0.0cm 12.cm 15.5cm,angle=90]{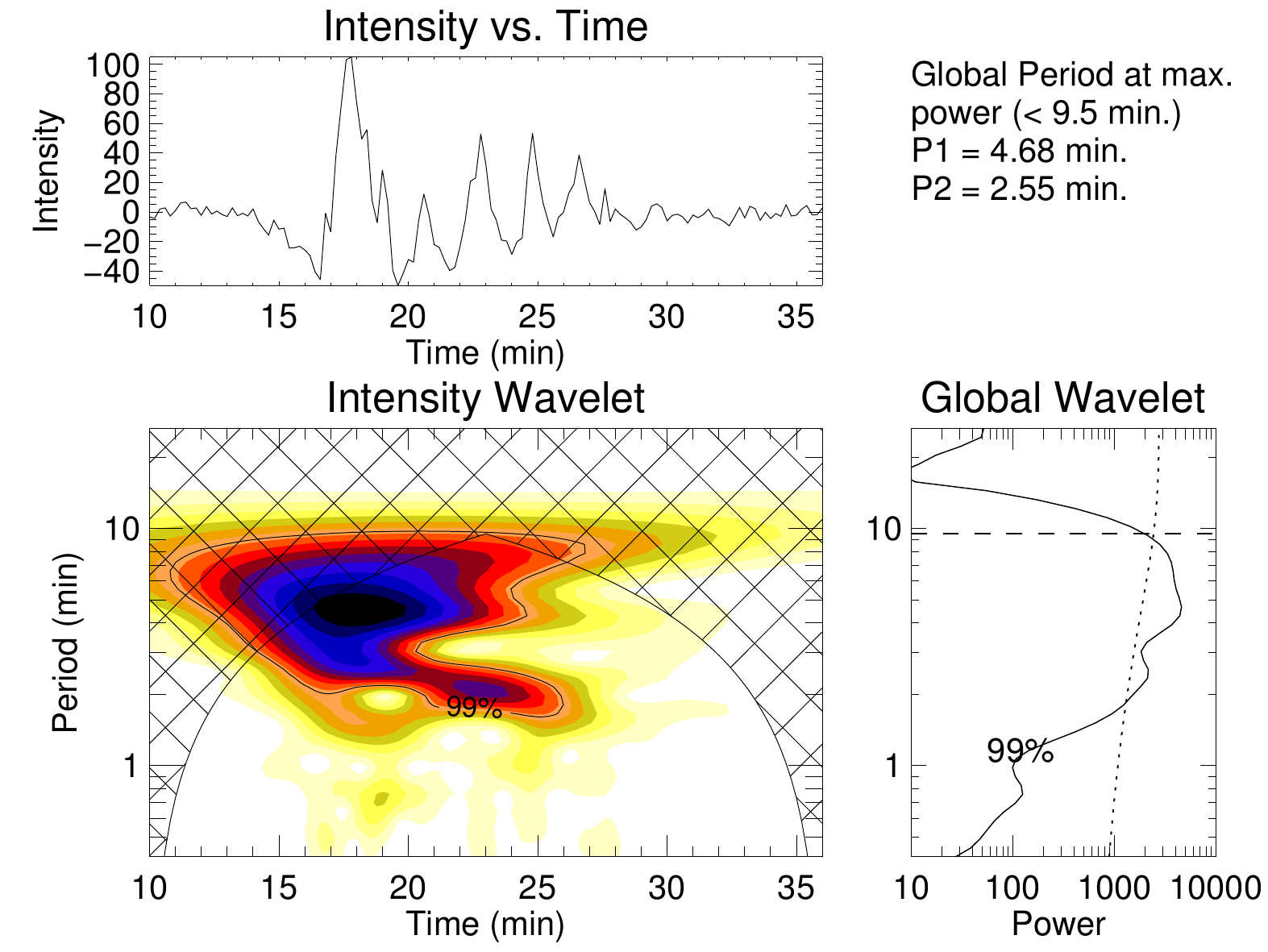}
\caption{The wavelet result for the quasi-periodic behaviour of the jet outflow events as observed from 
AIA 171~{\AA} channel. The top panels shows background trend removed (12-point running average) 
intensity variation with time. The bottom-left panel show the wavelet power 
spectrum with 99~\% confidence-level contours, whereas the bottom-right panel show the global wavelet 
power spectrum with 99~\% global confidence level. The periods P1 and P2 at the locations of the first 
two maxima in the global wavelet spectrum are also shown above the global wavelet spectrum. }
\label{fig:wavelet}
\end{figure}

Placing a 3x5 box along the main axis of the jet in 171~{\AA}, we obtain a light-curve averaged over the 
box. The running average of 12 frames is then subtracted from the curve and the de-trended data (top 
panel of Figure~\ref{fig:wavelet}) is subject to wavelet analysis (bottom panel of Figure~\ref{fig:wavelet}).
Details on the wavelet analysis, are described in \cite{TORCOMP1998} and provides information
on the time variation of intensity. In the wavelet spectrum, the cross-hatched regions are locations where 
estimates of the oscillation period becomes unreliable. The global wavelet plots are obtained by taking 
the mean over the wavelet time domain which makes it very similar to the Fourier transform (oscillatory 
power distribution with respect to period). The periods with first two strong oscillatory power peaks in 
the global wavelet spectrum are labelled above the global wavelet spectrum. The wavelet power spectra 
demonstrate the quasi-periodic behaviour of the outflow events, which has a
period of about $\sim150$~s. The presence of quasi-periodic outflows has previously been noted in 
\cite{MORetal2012b} for EUV jets and \cite{MOR2012} for chromospheric jets.


\section{Transverse oscillations on jet}
\label{sec:waves}

The jet shows signs of apparent wave motion in the fine-scale structure, which we now describe. The 
observed motion is a transverse displacement of the jet's axis. The $171$~{\AA} data needs to be further processed for wave studies. Firstly, the 
data is aligned using cross-correlation to achieve sub-pixel accuracy, with an apparent root mean 
squared frame-to-frame shift of $0.03$~pixels remaining. The signal to noise in the coronal hole is 
relatively low, so the data is then smoothed with a $3\times3$ boxcar function  to increase the signal to 
noise. We then apply the unsharp mask technique to reveal the fine-structure of the jet. The data noise is 
calculated using the formulae given in \cite{2012A&A...543A...9Y}, however, the boxcar smoothing  
reduces the data noise by a factor of a third. 

\begin{figure*}
\centering
\includegraphics[scale=.7]{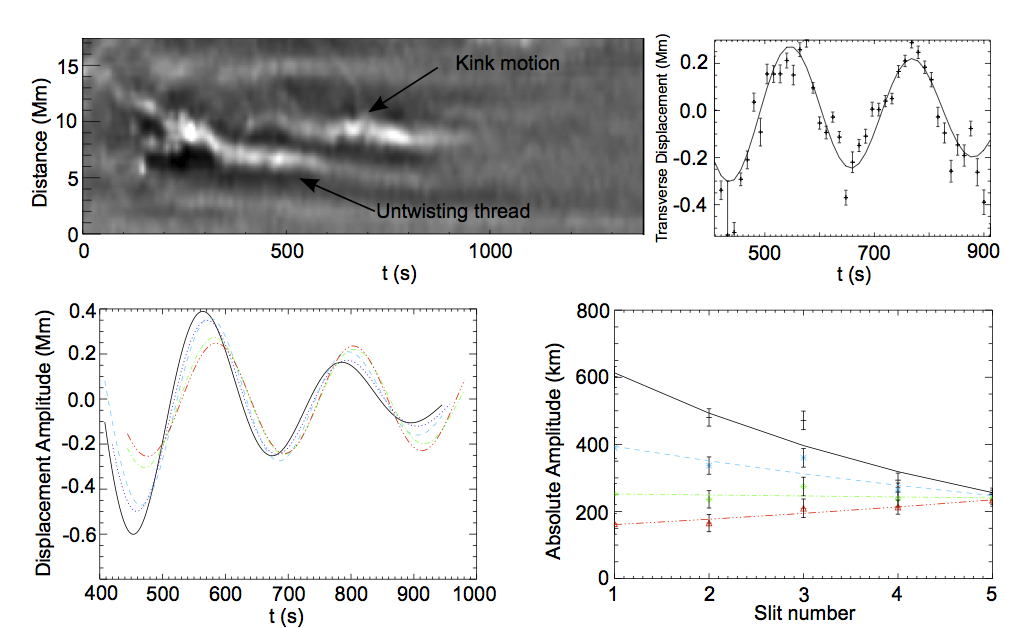}
\caption{The upper left panel shows a time-distance digram created using unsharp masked data. The 
time is seconds from 21:34:00~UT. Two neighbouring threads of the jet are clearly visible. The observed 
motion in the lower thread is likely due 
to the field line untwisting after reconnection, while the upper thread supports a transverse MHD wave. 
The upper right panel displays the results from the fitting process. The data points are obtained from a 
Gaussian fit and the error bars show the total error (the addition of the sigma errors derived from the 
Gaussian fit and the residual alignment errors). The solid line shows the sinusoidal fit to the data points. 
Note the data points have been de-trended by subtracting the G(t) term (see, Eq.~\ref{eq:fit}). The lower 
left panel shows the de-trended fits in all slits, slit 1 is solid/black, slit 2 is dotted/purple, slit 3 is 
dashed/blue, slit4 is dash-dot/green and slit 5 is dash-triple dot/red. The bottom right panel displays 
the decrease in amplitude as a function of slit. The amplitudes and sigma errors are plotted along with 
the exponential fits (lines) for the first minimum (black/crosses/solid), first maximum (dash/stars/blue), 
second minimum (dash-dot/diamonds/green) and second maximum (dash-triple dot/triangles/red).  }
\label{fig:wave}
\end{figure*}

From the filtered data time-distance diagrams are created, which reveal the transverse wave motion.  
Fig.~\ref{fig:wave} shows the cross-cut taken close to the foot-point of the jet. There are two readily 
identifiable features in the given time-distance diagrams. The longer duration event is a signature of the 
magnetic field untwisting as it rises into the solar atmosphere. The motion in time-distance diagrams 
looks distinctly like transverse wave motion but is probably best described by a helical or torsional 
motion. The shorter duration event is a transverse wave and has a large enough total displacement that it 
can be identified by eye in movies of the event. It should be pointed out that the second transverse waves 
occurs at a time ($\sim$21:42) when the intensity of the jet event has decreased significantly (see, 
Fig.~\ref{aia-lc}), hence the main violent and eruptive phase has finished, although the periodic 
out-flows are still occurring. The lower values of intensities would suggest the plasma along the jet spire 
is likely cooler and/or less dense when the wave phenomena occurs than during the first phase of plasma 
ejection.

The identification of the wave mode is aided by the observation of the physical displacement of the jet's 
axis. MHD wave theory demonstrates that in a cylindrical flux tube the fast magnetoacoustic kink mode is 
the only wave mode that perturbs the flux tube in such a manner. In the long wavelength limit, the kink 
mode has a phase speed equal to the kink speed \citep{1983SoPh...88..179E}:
\begin{equation}
 c_{k}=\sqrt{\frac{B_i^2+B_e^2}{\mu_{0}(\rho_{i}+\rho_{e})}} 
\end{equation}
where, $B$ the magnetic field strength, $\mu_{0}$ is the magnetic 
permeability of free space, $\rho$ is the density and the subscripts $i, e$ refer to the jet and ambient 
plasma quantities, respectively.

We measure the transverse wave by fitting a Gaussian function to the cross-sectional flux profile of the 
jet axis in each time slice (see, e.g., \citealp{2012NatCo...3E1315M}), providing the data noise as errors 
on the fit. The Gaussian centroid is taken as the central position of the jets axis, and to the associated 
errors on the fit we add the estimated remaining displacement error. These data points are then fitted with a damped sinusoidal function of the form, 
\begin{equation}\label{eq:fit}
F(t)=A_{0} \sin(\omega t+\phi)  \exp{(-\xi t)}+G(t),
\end{equation}
where  $A_0$ is the amplitude of the transverse displacement, $\omega=2\pi/P$ is the frequency and 
$P$ is the period, $\xi=1/{\tau}$ where $\tau$ is the damping time-scale and $G(t)$ is a linear function. 
The results of the wave fitting are shown in Fig.~\ref{fig:wave}. The velocity amplitude can be calculated 
using $v=2\pi A_0/P$.

We are able to fit the wave over 5 cross-cuts which are separated by approximately one pixel. The results 
for the wave fitting in each cross-cut are given in Table.~\ref{tab:kink}. It is apparent that the wave 
propagates as 
there is a delay between observing the wave in each of the cross-cuts. The arrival times of the maximum 
and minimum displacements are used to calculate the phase speed averaged across the 6 cross-cuts, 
which gives $v_{ph}=88\pm7$~km\,s$^{-1}$. Here, the 
$\pm7$~km\,s$^{-1}$ corresponds to the standard deviation of the measured velocities. Correcting for 
the jet's inclination the phase speed is $v_{ph}=110\pm10$~km\,s$^{-1}$.

The wave can also seen to be heavily damped as it propagates along the jet. The damping time-scale 
given from the fit parameters is not the damping time. Because the wave is propagating along the 
structure, the wave damping is a function distance, i.e., a damping length, $L_D$. 

The damping length can be calculated from the decrease in amplitude of the wave in each slit, fitting an 
exponential profile gives $L_D=1990\pm170$~km, $3740\pm770$~km, 
$3.6\times10^5\pm90\times10^5$~km and $-4610\pm1910$~km for the first minimum, first 
maximum, second minimum and second maximum, respectively. Correcting for the inclination of the jet 
with respect to the SDO projection gives $L_D=2470\pm210$~km, $4620\pm950$~km, 
$4.4\times10^5\pm1\times10^6$~km and $-5700\pm2360$~km. The increase in damping length is 
clear and there is even the suggestion of a slight amplification of the wave. 

{For damped oscillators, the size of the quality factor ($\xi_E$) determines whether the oscillation
is under damped, critically damped or over damped (e.g., \citealp{FRENCH}). The relation for the 
quality factor for propagating waves is
\begin{equation}\label{eq:qf}
\xi_E=\frac{L_D}{\lambda}=\frac{L_F f}{c_k},
\end{equation}
where $\lambda$ is the wavelength of the oscillation and $f=1/P$ is the frequency (note, this is different from the angular frequency 
$\omega$). The values of $\xi_E$ are then determined using the observationally derived values of 
phase speed, frequency and damping length; where it is found that $\xi_E$=0.10$\pm$0.09 and $0.19\pm0.04$ for 
$L_D=2470\pm170$~km and $4620\pm770$~km respectively.}

{
Critical damping occurs when $\xi_E=1/2\pi\approx0.16$, hence the results suggest that the first wave front is 
classed to lie in the over damped regime ($\xi_E<1/2\pi$). The consequence of this is that the first hundred seconds of the wave motion 
is evanescent in nature (\citealp{FRENCH}). For the remaining wave fronts, $\xi_E>1/2\pi$ and the wave motion is under damped and 
oscillatory.}

{
Observations of kink motions in coronal loops typically show a quality factor $>0.16$ and appear to be under damped oscillators 
(e.g., \citealp{VERetal2013}). Often, resonant absorption \citep{2002ApJ...577..475R,2002A&A...394L..39G} is cited as the mechanism
behind the observed damping, with the theoretical predictions providing a good agreement with the observations 
(e.g., \citealp{VERTetal2010}). If we assume that the observed damping is due to resonant absorption, then the quality factor is given by 
the TGV relationship derived by 
\cite{2010A&A...524A..23T}, i.e.,
\begin{equation}\label{eq:TVG}
L_D=v_{ph}\xi_E\frac{1}{f},
\end{equation}
which is identical to Eq.~\ref{eq:qf}. The magnitude of the quality factor is dependent on the thickness of the inhomogeneous layer, $l$,
between an over-dense flux tube and its environment, and the gradient of the density contrast, i.e.,
\begin{equation}\label{eq:TVG2}
\xi_E=\frac{2}{\pi}\frac{R}{l}\frac{\rho_i+\rho_e}{\rho_i-\rho_e}.
\end{equation}
Here, $R$ is the radius of the flux tube supporting the oscillation. Note that Eq.~\ref{eq:TVG2} is derived for weakly damped wave 
motion.} 

{
If resonant absorption is responsible for the observed damping, then the over damped motion suggest that the density contrast is large,
such that $\rho_e/\rho_i<<1$, or the inhomogeneous layer is thick. It is still unclear why the damping length changes for different 
cycles of the wave. It may be that as the jet develops, the internal density of the plasma along the jet spire tends towards the background 
coronal hole value. This would lead to a change in density ratio, i.e., $\rho_i/\rho_e\rightarrow1$, hence 
the damping length would increase. However, it may well be that other mechanisms are acting on the system, working together or individually, to bring about the enhanced damping, e.g., mode conversion/coupling, non-linearity (\citealp{OFM2009}; \citealp{RUDetal2010}).  } 

Note, the damping time-scale, $\tau$ represents the change in amplitude of the wave-packet that 
propagates through the slit, hence it demonstrates that the driver of the wave decreases with amplitude 
in time. The increase in the damping time-scale with distance along the jet axis is likely due to 
combination of decrease in driving amplitude with time, the wave being damped as a function of distance 
and the change in the jet parameters over time.

\begin{table*}
\caption{Measured parameters of the transverse waves.}
\centering
\begin{tabular}{c c c c c}
\hline \hline
Slit & Displacement & Period (s) & Velocity & Damping \\
& Amplitude (km) &  & Amplitude (km/s) & time (s)\\
\hline
1 &  $725 \pm 31 $ & $221\pm2$ & $20.6 \pm 0.9$ & $255\pm$14 \\
2 & $570 \pm 27 $ & $223\pm2$ & $16.1 \pm 0.9$  & $316\pm22$\\
3 & $544 \pm 28 $ & $220\pm1$ & $15.5 \pm 0.8$  & $408\pm35$ \\
4 & $ 312 \pm 21 $ & $222\pm2$ & $8.8 \pm 0.6$   & $1043\pm272 $ \\
5 & $ 256 \pm 18 $ & $218\pm2$ & $7.4 \pm 0.5$   & $4300\pm4300 $ \\
\hline
\end{tabular}\label{tab:kink}
\end{table*}

\section{DEM analysis}
Now, we estimate the density and temperature of the jet plasma from the emission measure of
the jet using the automated DEM tool described in \cite{2011SoPh..tmp..384A}. The DEM method 
assumes that the plasma is isothermal, in ionization equilibrium and has Maxwellian velocities. It is 
unclear whether all these conditions are satisfied by the jet plasma. We discuss the potential influence of 
these effects at the end of this section.  

In Figure~\ref{171dem} we show the results of the DEM analysis. The top panel shows a zoomed portion 
of the jet plasma with the axes in units of solar radii. The jet length is divided in to 45 spatial bins for 
the calculation of the DEM. The bottom three panels show the electron temperature, number densities 
and width along the length of the jet. We estimated the electron density and temperature of the 
jet material for the time sequence starting from 21:39:12 UT to 21:40:36 UT, for nine consecutive frames 
in order to assess how rapidly plasma parameters evolve in the jet. Averaged values of the electron 
number density and temperature corresponding to each time sequence selected is given in 
Table~\ref{T2}. In general, the DEM shows relatively consistent values of $n_e$ and $T$. Electron number density and temperature of the 
plasma outside the jet cannot be 
calculated since the method is not suitable for DEM measurements for uniform density background 
plasma. For calculations in the following section we take a value of log ($n_{e}) = 8.1$, as a reference 
value for the number density corresponding to the ambient coronal hole
 (\citealp{1998A&A...339..208B}; \citealp{CRA2009}). {In \cite{1998A&A...339..208B}, the electron density was estimated from 
the intensity line ratios offorbidden spectral lines of \ion{Si}{viii} and CHIANTI database. The spectra were obtained
with the Solar Ultraviolet Measurements of Emitted Radiation spectrometer (SUMER) flown on the Solar and Heliospheric Observatory
spacecraft. The relative measuring error in deriving $n_e$ is estimated to be 12\% to 15\%, similar to the expected absolute error (see
\citealp{LAMETAL1997}). \cite{DOYETAL1998} measured electron densities at the equatorial limb, again using the \ion{Si}{viii}
line ratio. Comparing their Table 1 with the \cite{1998A&A...339..208B}, the densities in the coronal holes are significantly lower than the
‘quite’ Sun, approximately a factor of two, as also noted by \cite{DOSetal1977}.}

 
 \begin{table}
\caption{Estimates of the average electron number density along the length of the jet. }
\label{T2}
\centering
\begin{tabular}{ccc}
\hline 
Time in seconds & Density & Temperature \\
starting from &  log($n_{e}$) & log($T_e$)  \\
 
 22:39 UT&  \\
 
\hline
\\

12 & 8.75$\pm$0.05 & 5.89$\pm$0.08 \\
24 & 8.85$\pm$0.10 & 5.79$\pm$0.07 \\
36 & 9.61$\pm$0.25 & 6.46$\pm$0.46 \\
48 & 8.62$\pm$0.04 & 5.84$\pm$0.05 \\
60 & 8.82$\pm$0.08 & 5.73$\pm$0.06 \\
72 & 8.75$\pm$0.11 & 5.79$\pm$0.07 \\
84 & 8.80$\pm$0.12 & 5.74$\pm$0.08 \\
96 & 8.81$\pm$0.09 & 5.71$\pm$0.03 \\

\hline 

\end{tabular}
\end{table}

The temperatures given in Table~\ref{T2} suggest the jet is relatively cool, i.e., compared to jets emitting 
in X-ray wavelengths ($T\sim2$~MK). However, we believe the measured value is supported by the STEREO data. In 
particular, the jet is seen clearly in 304~{\AA} but is not distinguishable in 195~{\AA} from the emission 
along the line of sight. This can be explained as follows. The jet is clearly visible in SDO 193~{\AA} which 
has a similar temperature coverage as STEREO 195~{\AA} (e.g., \citealp{HOWetal2008}). However, SDO is 
able to view the jet in the coronal hole with little surrounding hot ($T>1$~MK) plasma. In contrast, there 
exists closed coronal structures along the line-of-sight between the jet and STEREO A/B viewpoints 
which strongly contribute to the observed emission in 193/195~{\AA} passbands. Hence, the 
relatively cooler temperature (in comparison to the closed corona) and approximate coronal densities of 
the jet means the jet's emission is not strong enough to stand out amongst the general coronal emission
(as viewed by STEREO).

Now, it is worth discussing the potential fragilities associated with the DEM results. Concerning ionization 
equilibrium, rapid injections of heat can drive the plasma into a state of non-equilibrium ionisation (e.g., 
\citealp{BRAMAS2003}, \citealp{BRAetal2004}), leading to ions of relatively low charge existing at higher 
temperatures than usual. The temperature estimate obtained from the DEM , $T\approx10^{5.89}$~K, 
suggests the plasma temperature is close to the peak formation temperature of \ion{Fe}{ix} (i.e., 
$T\sim10^{5.8}$~K), a dominant contributor to the observed emission in 171~{\AA}. If 
the jet plasma has been heated rapidly from a cooler temperature, it will lead to delay in ionisation rates  
and have influence on emission in 171~{\AA}. Hence, in this scenario, the DEM technique would 
underestimate the plasma temperature. Further, the electron density, which is calculated from derived 
value of EM, would also be underestimated, along with the plasma density (if using the typical coronal 
assumption that the plasma is fully ionised, i.e.,  $n_e\approx n_i$).

However, the time-scales for ionisation equilibrium are dependent on the heating rate and it is calculated 
that equilibrium is reached in matter of seconds to minutes (e.g., \citealp{BRAMAS2003}, 
\citealp{BRAetal2004}). \cite{DEMBRA2008} also demonstrate the quasi-periodic heating events (e.g., 
periodic reconnection) could lead to the plasma being in a state of constant non-equilibrium. For this to 
occur, the time-scales for heating would have to be of a relatively high-frequency. If the multiple, periodic 
ejections observed in the jet do correspond to heating events, the large periodicity ($\sim150$~s) means 
the jet plasma has time to reach ionisation equilibrium.

Impulsive heating events may also accelerate particles to non-thermal (i.e., non-Maxwellian) velocities, 
which can influence ionisation rates (e.g, \citealp{DZI2006}, \citealp{DZIetal2011}). It is still unclear 
whether such situations occur in jet events, but should be kept in mind when interpreting the DEM 
results.
 
\section{Magneto-seismology}
Now, we combine the results from the wave fitting and the DEM analysis to estimate the magnetic field 
strength along the jet spire. The average magnetic field in the jet 
material can be estimated from the definition of the kink speed and is given by,

\begin{equation}
 B =\frac{c_{k}}{\sqrt{2}} \sqrt{\mu_0(\rho_{e}+\rho_{i})  },
\end{equation}
where $B$ is the average magnetic field given by $B^2=(B_i^2+B_e^2)/2$. First, we estimate the plasma 
density from the electron number density. Taking a typical value for the electron density from Table~2, 
i.e., $\log{8.8\pm0.1}$~cm$^{-3}$, the plasma density for a fully ionized plasma can be estimated using 
$\rho=\tilde{\mu} m_Hn_e$, where $m_H=1.67\times10^{-27}$~kg is the mass of hydrogen and 
$\tilde{\mu}\approx1.27$ is the mean molecular weight per atom. Hence, the internal and 
external plasma densities are $\rho_i=1.69\pm0.43\times10^{-12}$~kg\,m$^{-3}$ and 
$\rho_e=2.5\pm0.4\times10^{-13}$~kg\,m$^{-3}$ . {Using the measured value of kink speed ($c_k=110\pm10$~km\,s$^{-1}$)}, the magnetic field of the jet spire is estimated to be $B=1.21\pm0.2$~G (corrected for 
inclination). To calculate the error on the magnetic field strength we used the following formulae
\begin{equation}
\delta B=\sqrt{\frac{\mu_0}{2}}\left\{(\rho_i+\rho_e)\delta c_k^2+\frac{c_k^2(\delta\rho_i^2+\delta\rho_e^2)}
{4(\rho_i+\rho_e)} \right\}^{1/2},
\end{equation}
and it is assumed that the error on the external density is value is equal to the error on the internal 
density value. The $\delta x$ corresponds to the associated error in the parameter $x$.

In order to compare these derived values of field strength to previous measurements and 
expectations of coronal hole magnetic fields, it is necessary to mention the typical magnetic geometry of 
coronal holes. From the correlation of Doppler velocities in a number of different spectral lines 
(\citealp{TUetal2005b}, \citealp{TIAetal2008}) that are assumed to be formed at various heights in the 
atmosphere, the results appears to show that coronal holes are composed of rapidly expanding magnetic 
funnels. To date, the majority of measurements of magnetic field strength in coronal holes have 
been of the underlying, photospheric field and then extraplotaions have provided coronal values 
(\citealp{TUetal2005,TUetal2005b}; \citealp{WAN2010}; \citealp{HEetal2010b}). Magnetic 
field strengths at the photosphere of coronal 
holes seem to be $\mid B\mid<200$~G (e.g., \citealp{WAN2010}) with average values of unsigned flux 
are $<\mid B\mid>\sim1-40$~G (see, e.g., \citealp{WEISOL2004} and references therein).

Using HMI data from a $50''\times 40''$ region centred on the jet, we find the largest value of field 
strength reaches $\sim-250$~G (line-of-sight value $\sim200~G$) with an average unsigned value of 
$\sim11$~G (line-of-sight value $\sim9~G$). To compare this to our estimated coronal value, we follow \cite{HEetal2008} and assume the variation of magnetic field strength with height for the 
magnetic funnels is exponential, i.e.,
\begin{equation}
B(z)=b_0\exp\left(-\frac{z}{H_B}\right).
\end{equation} 
Here, $b_0$ is the photospheric field value and $H_B$ is the magnetic scale height, which we take to be
$3.57$~Mm (\citealp{TUetal2005b}). Using the average of value of unsigned magnetic flux we obtain a 
value of $B(z)\sim1.2$~G at a height of $z\approx7.9$~Mm. Hence, this value is consistent with the 
height of the base of the corona at which the transverse wave is measured.

Recently, \citet{CHEetal2012} also attempted to estimate the magnetic field strength of a solar jet in 
polar coronal hole, suggesting at field strength of 15 G at 11~Mm that decreased to around 3 G at 
28~Mm. However, the authors use a mean photospheric value of magnetic flux density from active region 
observations (i.e., $\sim500$~G). If they would have used typical measured coronal hole values for the 
average field strength, $\sim10$~G, the value of field strength at 11~Mm would be $0.3$~G, which is 
consistent with the estimate of weak magnetic field strength that we obtain. 

In different (possibly open-field) regions of the solar atmosphere, \citet{2011ApJ...730..122W} performed 
a seismological study using an EIT waves and estimated coronal magnetic field strength to be 
relatively weak ($0.7\pm 0.7$~G) in the quiet Sun.

{
\begin{figure}
\centering
{\includegraphics[width=8cm, bb=290 482 553 748,clip=]{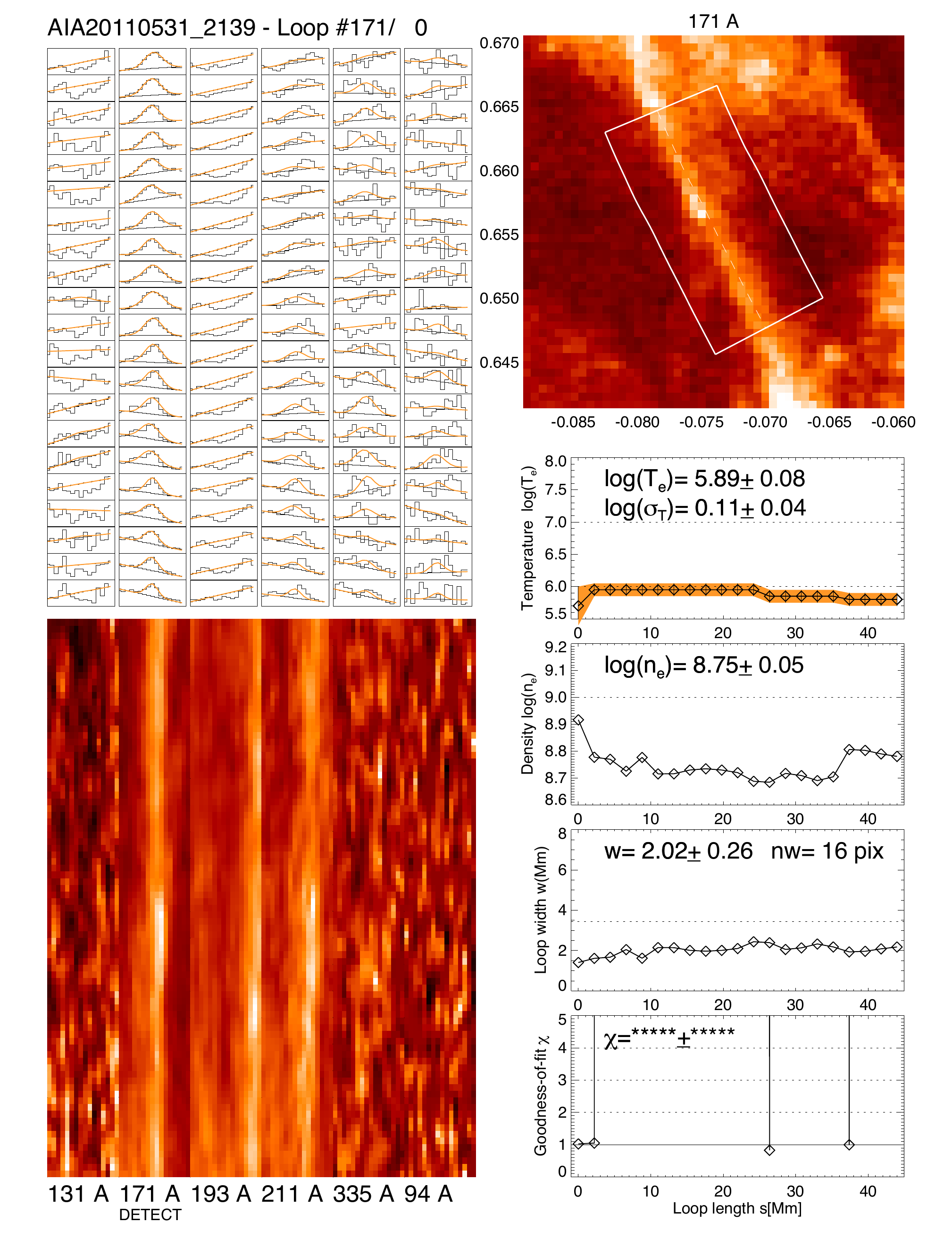}} \\
{\includegraphics[width=8cm,bb=292 252 553 480,clip=]{t3912}}
\caption{Example result from DEM analysis. Top panel shows the selected section of the jet plasma as 
seen in the 171~{\AA} line used for the emission-measure and temperature analysis (axes are in solar 
radii units). The middle panel is the estimated electron temperature along the jet, with the horizontal axis 
corresponding to bin number along the jet. Bottom panel shows the variation of electron number density 
along the length of the jet.}
\label{171dem}
\end{figure}

\section{Conclusion}
\label{sec:discussion}
We present here the observation and an analysis of a jet in a solar coronal hole, close to the holes 
boundary. The observed features during the evolution of the jet are discussed with respect to the 
knowledge of jet formation from numerical simulations (e.g., \citealp{MORGAL2013}). We interpret 
enhanced emission in multiple AIA bandpasses as evidence for the jet being initiated by the emergence 
of new flux into the region and the formation of current sheets and sites of heating during the initial stage of the 
jet's formation. However, the HMI data does not appear to contain evidence for emerging flux. We can 
only suggest that the emerging flux is relatively weak. 

Following the initial phase, the jet spire is formed and is associated with fast out-flowing plasma 
($>$360~km\,s$^{-1}$). After an initial ejection, there is a delay before multiple smaller, quasi-periodic 
($\sim150$~s) outflows occur along the spire. This may be related to oscillatory reconnection events or 
even too multiple eruptive phases due to various instabilities of the system (see, e.g., 
\citealp{MORGAL2013}).

During the latter stages of the jet's evolution, we are able to identify a transverse oscillatory motion of 
the jet's spire. Using an advanced fitting technique we are able to measure the properties of the 
transverse displacement with a high degree of accuracy. The wave is found to propagate along the spire 
with speeds of 110~km\,s$^{-1}$. In addition, we exploit the latest differential emission measure 
techniques to estimate the temperature 
and electron density along the jet spire, obtaining values of $\log T\sim 5.8$~K and $\log 
n_e\sim8.8$~cm$^{-3}$. These values are in agreement with previous measured values using a variety of 
techniques (e.g., \citealp{DOSetal2010}; \citealp{NISetal2011}). We then demonstrate that, using the 
measured values of the observed transverse wave and the density estimates from the DEM, we are able to 
obtain an estimate for the magnetic field strength ($B=1.21\pm0.2$~G ) of the section of the jet structure 
that supports the wave. 

In conclusion, we demonstrate the potential of combining magneto-seismology with spectroscopic 
techniques for obtaining insights into local plasma parameters. This work provides support for previous 
measurements of density and temperature using spectrometers. However, by combining the two 
techniques we are able to reveal additional information on the magnetic field in the coronal hole. 
Future studies that obtain detailed information on the plasma parameters of particular jet events will 
provide valuable constraints for future modelling efforts.

\acknowledgements{The authors thank M.~Aschwanden and G.~Verth for enlightening discussions and the anonymous referee for
their comments that improved the manuscript. The 
authors would also like to thank the Royal Society for support that made this work possible. This work is 
supported by the UK Science and Technology Facilities Council (STFC), 
with RM grateful to the Northumbria University for the award of the Anniversary Fellowship.  }

\bibliographystyle{aa}

\end{document}